\documentclass{article}

\input{tcilatex}

\begin{document}

\title{A simpler proof of zero-knowledge against quantum attacks using
Grover's amplitude amplification}
\author{Keiji Matsumoto}
\maketitle

\section{Introduction}

Watrous\cite{Watrous2005} had presented the first proof of zero-knowledge
property of a proof system against a quantum verifier. The key of the proof
is the construction of a quantum simulator. In the construction, the
'failure state' is rotated to the 'success' state by a tricky operation
which is initially developped for the amplification of QMA proof systems.

This manuscript presents a new and simpler construction of a simulator. In
the construction, we simply amplify the success probability of a classical
simulator using Grover's amplification.

\section{The Goldreich-Micali-Wigderson Graph Isomorphism Proof System}

The Goldreich-Micali-Wigderson graph isomorphism protocol is a well-known
example of a proof system that is perfect zero-knowledge against classical
polynomial-time verifiers. In this section it is proved that this protocol
is in fact zero-knowledge against polynomial-time quantum verifiers. The
method can be extended to several other protocols.

\subsection{The protocol}

Let $x$ be a pair of graphs $(G_{0},G_{1})$, and $L$ be a set of pairs with $%
G_{0}\simeq G_{1}$. Hereafter, $P$ denotes the prover, and $V$ the verifier.

\begin{description}
\item[(a)] $P$ randomly chooses a permutation $\tau $ on the graph, and
sends $\tau (G_{0})$ to $V$.

\item[(b)] $V$ sends a random bit $a\in \{0,1\}$ to $P$.

\item[(c)] $P$ send a permutation $\pi $, and $V$ accepts if $\tau
(G_{0})=\pi (G_{a})$.
\end{description}

To decrease the error probability, (a)-(c) are repeated for polynomially
many times.

The quantum description of this classical protocol is as follows. Let $%
\mathcal{V}$ and $\mathcal{A}$ be the $V$'s workspace and a qubit which
stores output of the simulator at the end the step (b), respectively. The
register $\mathcal{Y}$ stores the message from $P$ to $V$ in the step (a).
We also denote by $\mathcal{W}$ the register for an auxiliary input $%
\left\vert \psi \right\rangle $.

The initial state is%
\[
\left\vert \psi \right\rangle |0_{\mathcal{V}}\rangle |0_{\mathcal{A}%
}\rangle |0_{\mathcal{Y}}\rangle . 
\]%
After the step (a),%
\[
\left\vert \psi \right\rangle \left\langle \psi \right\vert \otimes |0_{%
\mathcal{V}}\rangle \left\langle 0_{\mathcal{V}}\right\vert \otimes |0_{%
\mathcal{A}}\rangle \left\langle 0_{\mathcal{A}}\right\vert \otimes \frac{1}{%
n!}\sum_{\tau \in S_{n}}|\tau (G_{0})\rangle \left\langle \tau
(G_{0})\right\vert 
\]%
The honest verifier will apply Hadamard transform to $|0_{\mathcal{A}%
}\rangle $ and measure $\mathcal{A}$ in the step (b),%
\[
\frac{1}{2n!}\left\vert \psi \right\rangle \left\langle \psi \right\vert
\otimes |0_{\mathcal{V}}\rangle \left\langle 0_{\mathcal{V}}\right\vert
\otimes \sum_{a\in \{0,1\}}|a_{\mathcal{A}}\rangle \left\langle a_{\mathcal{A%
}}\right\vert \otimes \sum_{\tau \in S_{n}}|\tau (G_{0})\rangle \left\langle
\tau (G_{0})\right\vert . 
\]%
In general, however, a verifier will apply an unitary transform $U_{V}$ on $%
\mathcal{W\otimes V\otimes A\otimes Y}$, and measure $\mathcal{A}$.%
\[
\frac{1}{n!}\sum_{\tau \in S_{n},a\in \{0,1\}}|a_{\mathcal{A}}\rangle
\left\langle a_{\mathcal{A}}\right\vert U_{V}\left( \left\vert \psi
\right\rangle \left\langle \psi \right\vert \otimes |0_{\mathcal{V}}\rangle
\left\langle 0_{\mathcal{V}}\right\vert \otimes |0_{\mathcal{A}}\rangle
\left\langle 0_{\mathcal{A}}\right\vert \otimes |\tau (G_{0})\rangle
\left\langle \tau (G_{0})\right\vert \right) U_{V}^{\dagger }|a_{\mathcal{A}%
}\rangle \left\langle a_{\mathcal{A}}\right\vert . 
\]%
After this, the step (c) follows, but we omit the description of this part,
for this step is easy to simulate once a simulation of the steps (a)-(b) is
given.

\subsection{A simulator}

A classical simulator is constructed as follows. Assume that $G_{0}\simeq
G_{1}$. The simulator randomly chooses $b\in \{0,1\}$ and $\pi \in S_{n}$,
and compute $\pi (G_{b})$ which mimics $P$'s first message. Then it applies
the operation of $V$ on the simulated message from $P$, producing an output $%
a\in \{0,1\}$, or the message to $P$. If $a=b$, $\pi $ chosen previously can
mimic the second message from $P$ to $V$, and the simulation succeeds. If $%
a\neq b$, we "rewind", or abort and restart from the beginning. This
successfully simulates the single round of GMW protocol with probability $%
\frac{1}{2}$, meaning that the simulation succeeds with high probability
after some iterations.

To simulate the iterations of the single round, the simulator also has to be
repeatedly run. Observe that in rewinding, the simulation only has to
restart from the beginning of the present round, with the record of the
final state of the previous round being copied in some registers. Otherwise,
the simulation would take exponential time. In quantum case, however, this
part fails because of the no-cloning principle.

Here we show how to bypass this difficulty: Grover's amplitude amplification
can increase the success probability of the simulation of each round up to
1, and thus there is no need for rewinding.

Let us define 
\[
\mathcal{X=V\otimes A\otimes Y\otimes B\otimes Z}, 
\]%
where $\mathcal{Z}$ and $\mathcal{B}$ stores random bits specifying a
permutation $\pi $ on the graph and a random bit $b$, respectively.

Let us denote by $A$ a unitary operation corresponding to the the classical
simulator other than rewinding part, \vspace{-1mm}%
\begin{eqnarray*}
&&A\left\vert \psi \right\rangle \left\vert 0_{\mathcal{X}}\right\rangle = \\
&&\frac{1}{\sqrt{2n!}}\sum_{b\in \{0,1\},\pi \in S_{n}}\left(
U_{V}\left\vert \psi \right\rangle \left\vert 0_{\mathcal{V}}\right\rangle
\left\vert 0\right\rangle \left\vert \pi (G_{b})\right\rangle \,\right)
\left\vert b\right\rangle \left\vert \pi \right\rangle .
\end{eqnarray*}%
We apply amplitude amplification to this operation. Define a unitary
transform $S_{0}^{\phi }$, $S_{1}^{\varphi }$ in $\mathcal{W}\otimes 
\mathcal{X}$ by 
\begin{eqnarray*}
S_{0}^{\phi } &:&=\left( \phi -1\right) \mathbf{I}_{\mathcal{W}}\otimes
\left\vert 0_{\mathcal{X}}\right\rangle \langle 0_{\mathcal{X}}|+\mathbf{I},
\\
S_{1}^{\varphi } &:&=\left( \varphi -1\right) \Pi ^{\lambda }+\mathbf{I}.
\end{eqnarray*}%
where $\Pi $ is the projection onto success event, 
\[
\Pi :=\sum_{b\in \{0,1\}}\mathbf{I}_{\mathcal{W}\otimes \mathcal{V}}\otimes
\left\vert b\right\rangle \left\langle b\right\vert \otimes \mathbf{I}_{%
\mathcal{Y}}\otimes \left\vert b\right\rangle \left\langle b\right\vert
\otimes \mathbf{I}_{\mathcal{Z}}. 
\]%
These phase factors are chosen according to lemma 3 in \cite{BHT98}.

Observe that $a=b$ occurs with probability $\frac{1}{2}$, for all the state $%
|\psi \rangle $ because $b\in \{0,1\}$ is uniformly random, and does not
affect the input of $U_{V}$. This assures us the identity 
\begin{equation}
\langle 0_{\mathcal{X}}|A^{-1}\Pi A|0_{\mathcal{X}}\rangle =\frac{1}{2}%
\mathbf{I}_{\mathcal{W}}.  \label{eq:P=1/2}
\end{equation}%
More rigorously, this is true for the following equalities holds for any $%
\left\vert \psi \right\rangle $:%
\begin{eqnarray*}
&&\left\Vert \Pi A\left\vert \psi \right\rangle \left\vert 0_{\mathcal{X}%
}\right\rangle \right\Vert ^{2} \\
&=&\frac{1}{2n!}\left\Vert \sum_{a,b\in \{0,1\},\pi \in S_{n}}\mathbf{I}_{%
\mathcal{W}\otimes \mathcal{V}}\otimes \left\vert a\right\rangle
\left\langle a\right\vert \otimes \mathbf{I}_{\mathcal{Y}}\otimes \left\vert
a\right\rangle \left\langle a\right\vert \otimes \mathbf{I}_{\mathcal{Z}%
}\left( U_{V}\left\vert \psi \right\rangle \left\vert 0_{\mathcal{V}%
}\right\rangle \left\vert 0\right\rangle \left\vert \pi (G_{b})\right\rangle
\,\right) \left\vert b\right\rangle \left\vert \pi \right\rangle \right\Vert
^{2} \\
&=&\frac{1}{2n!}\left\Vert \sum_{b\in \{0,1\},\pi \in S_{n}}\mathbf{I}_{%
\mathcal{W}\otimes \mathcal{V}}\otimes \left\vert b\right\rangle
\left\langle b\right\vert \otimes \mathbf{I}_{\mathcal{Y\otimes B\otimes Z}%
}\left( U_{V}\left\vert \psi \right\rangle \left\vert 0_{\mathcal{V}%
}\right\rangle \left\vert 0\right\rangle \left\vert \pi (G_{b})\right\rangle
\,\right) \left\vert b\right\rangle \left\vert \pi \right\rangle \right\Vert
^{2} \\
&=&\frac{1}{2n!}\sum_{b\in \{0,1\}}\sum_{\pi \in S_{n}}\left\Vert \mathbf{I}%
_{\mathcal{W}\otimes \mathcal{V}}\otimes \left\langle b\right\vert \otimes 
\mathbf{I}_{\mathcal{Y}}\left( U_{V}\left\vert \psi \right\rangle \left\vert
0_{\mathcal{V}}\right\rangle \left\vert 0\right\rangle \left\vert \pi
(G_{b})\right\rangle \,\right) \right\Vert ^{2} \\
&=&\frac{1}{2n!}\sum_{b\in \{0,1\}}\sum_{\pi \in S_{n}}\left\Vert \mathbf{I}%
_{\mathcal{W}\otimes \mathcal{V}}\otimes \left\langle b\right\vert \otimes 
\mathbf{I}_{\mathcal{Y}}\left( U_{V}\left\vert \psi \right\rangle \left\vert
0_{\mathcal{V}}\right\rangle \left\vert 0\right\rangle \left\vert \pi \tau
^{b}(G_{0})\right\rangle \,\right) \right\Vert ^{2} \\
&=&\frac{1}{2n!}\sum_{\pi \in S_{n}}\sum_{b\in \{0,1\}}\left\Vert \mathbf{I}%
_{\mathcal{W}\otimes \mathcal{V}}\otimes \left\langle b\right\vert \otimes 
\mathbf{I}_{\mathcal{Y}}\left( U_{V}\left\vert \psi \right\rangle \left\vert
0_{\mathcal{V}}\right\rangle \left\vert 0\right\rangle \left\vert \pi
(G_{0})\right\rangle \,\right) \right\Vert ^{2} \\
&=&\frac{1}{2n!}\sum_{\pi \in S_{n}}1=\frac{1}{2},
\end{eqnarray*}%
where in the third line, $\tau (G_{0})=G_{1}$. Using the equation (\ref%
{eq:P=1/2}), as shortly described, we can explicitely check the following
identity 
\begin{equation}
AS_{0}^{\imath }A^{-1}S_{1}^{\imath }A\left\vert \psi \right\rangle
\left\vert 0_{\mathcal{X}}\right\rangle =(\imath -1)\Pi A|\psi \rangle
\left\vert 0_{\mathcal{X}}\right\rangle .  \label{eq:GMW-amplify}
\end{equation}%
Measure $\mathcal{B}$ and $\mathcal{Z}$, and compute $\pi (G_{b})$, and
store its result some register, say $\mathcal{Z}^{\prime }$. Trace out the
register. Then, the final state is%
\begin{eqnarray*}
&&\frac{1}{n!}\sum_{\pi \in S_{n},a\in \{0,1\}}|a_{\mathcal{A}}\rangle
\left\langle a_{\mathcal{A}}\right\vert U_{V}\left( \left\vert \psi
\right\rangle \left\langle \psi \right\vert \otimes |0_{\mathcal{V}}\rangle
\left\langle 0_{\mathcal{V}}\right\vert \otimes |0_{\mathcal{A}}\rangle
\left\langle 0_{\mathcal{A}}\right\vert \otimes |\pi (G_{a})\rangle
\left\langle \pi (G_{a})\right\vert \right) U_{V}^{\dagger }|a_{\mathcal{A}%
}\rangle \left\langle a_{\mathcal{A}}\right\vert \\
&&\otimes |\pi (G_{a})\rangle _{\mathcal{Z}^{\prime }}\,_{\mathcal{Z}%
^{\prime }}\left\langle \pi (G_{a})\right\vert \\
&=&\frac{1}{n!}\sum_{\tau \in S_{n},a\in \{0,1\}}|a_{\mathcal{A}}\rangle
\left\langle a_{\mathcal{A}}\right\vert U_{V}\left( \left\vert \psi
\right\rangle \left\langle \psi \right\vert \otimes |0_{\mathcal{V}}\rangle
\left\langle 0_{\mathcal{V}}\right\vert \otimes |0_{\mathcal{A}}\rangle
\left\langle 0_{\mathcal{A}}\right\vert \otimes |\tau (G_{0})\rangle
\left\langle \tau (G_{0})\right\vert \right) U_{V}^{\dagger }|a_{\mathcal{A}%
}\rangle \left\langle a_{\mathcal{A}}\right\vert \\
&&\otimes |\tau (G_{0})\rangle _{\mathcal{Z}^{\prime }}\,_{\mathcal{Z}%
^{\prime }}\left\langle \tau (G_{0})\right\vert .
\end{eqnarray*}%
This shows that $\pi (G_{b})$, $\mathcal{W\otimes V\otimes A\otimes Y}$, and 
$\mathcal{Z}$ mimics the message from $P$ to $V$ in the step (a), the $V$'s
final state in the step (b) and the message from $V$ to $P$, and the message
from $P$ to $V$ in the step (c), respectively.

Below, we use the block representation in which $\left\vert \psi
\right\rangle \left\vert 0_{\mathcal{X}}\right\rangle $ writes 
\[
\left\vert \psi \right\rangle \left\vert 0_{\mathcal{X}}\right\rangle =\left[
\begin{array}{c}
|\psi \rangle \\ 
0%
\end{array}%
\right] . 
\]%
In that representation,%
\begin{eqnarray*}
A^{-1}\Pi A &=&\left[ 
\begin{array}{cc}
\frac{1}{2}\mathbf{I}_{\mathcal{W}} & \Pi _{A,12}^{\dagger } \\ 
\Pi _{A,12} & \mathbf{\ast }%
\end{array}%
\right] , \\
S_{0}^{\imath } &=&\left[ 
\begin{array}{cc}
\imath \mathbf{I}_{\mathcal{W}} & 0 \\ 
0 & \mathbf{I}_{\mathcal{X}}%
\end{array}%
\right] .
\end{eqnarray*}

Therefore,%
\begin{eqnarray*}
&&AS_{1}^{\imath }\cdot A^{-1}S_{0}^{\imath }A\left\vert \psi \right\rangle
\left\vert 0_{\mathcal{X}}\right\rangle \\
&=&AS_{1}^{\imath }\cdot \left( \left( \imath -1\right) A^{-1}\,\Pi A+%
\mathbf{I}\right) \left\vert \psi \right\rangle \left\vert 0_{\mathcal{X}%
}\right\rangle \\
&=&AS_{1}^{\imath }\left[ 
\begin{array}{c}
\left( \frac{\imath -1}{2}+\mathbf{1}\right) |\psi \rangle \\ 
\left( \imath -1\right) P_{A,12}|\psi \rangle%
\end{array}%
\right] \\
&=&A\left[ 
\begin{array}{c}
\imath \left( \frac{\imath -1}{2}+\mathbf{1}\right) |\psi \rangle \\ 
\left( \imath -1\right) \Pi _{A,12}|\psi \rangle%
\end{array}%
\right] \\
&=&(\imath -1)A\left[ 
\begin{array}{c}
\frac{1}{2}|\psi \rangle \\ 
\Pi _{A,12}|\psi \rangle%
\end{array}%
\right] \\
&=&(\imath -1)AA^{-1}\Pi A|\psi \rangle \left\vert 0_{\mathcal{X}%
}\right\rangle \\
&=&(\imath -1)\Pi A|\psi \rangle \left\vert 0_{\mathcal{X}}\right\rangle .
\end{eqnarray*}%
This is our assertion (\ref{eq:GMW-amplify}).

\subsection{Watrous's simulator revisited}

Instead of doing Grover's amplitude amplification, we can perform the
measurement $\Pi $ to the state $A|\psi \rangle |0_{\mathcal{X}}\rangle $.
If the success event is observed, we are done. This occurs with probability $%
\frac{1}{2}$. Otherwise, the state of the system colllapses to $\sqrt{2}(%
\mathbf{I}-\Pi )A|\psi \rangle |0_{\mathcal{X}}\rangle $, and $%
AS_{0}^{-1}A^{-1}$, or reflection about $A|\psi \rangle |0_{\mathcal{X}%
}\rangle $ maps this state to $\sqrt{2}\Pi A|\psi \rangle |0_{\mathcal{X}%
}\rangle $, which corresponds to success. 

This simulation is the same as the one presented in \cite{Watrous2005},
although the presentation is different.

\section{When success probability is not $\frac{1}{2}$}

\subsection{Amplification operations}

The construction in the previous section seemingly depends on the fact that
the success probability equals $\frac{1}{2}$. In the section, we show that
if we have

\[
A^{-1}\Pi A=\left[ 
\begin{array}{cc}
\lambda \mathbf{I}_{\mathcal{W}} & \Pi _{A,12}^{\dagger } \\ 
\Pi _{A,12} & \Pi _{A,22}%
\end{array}%
\right] 
\]%
our method works for any success probability $\lambda $, if proper phase
shifts are introduced. Especially, we have to check that repetition of the
amplification works in the same as the case where the auxiliary input $%
\left\vert \psi \right\rangle $ is absent.

Then, the identity%
\begin{eqnarray*}
&&\left( A^{-1}\Pi A\right) ^{2} \\
&=&\left[ 
\begin{array}{cc}
\lambda ^{2}\mathbf{I}_{\mathcal{W}}+\Pi _{A,12}^{\dagger }\Pi _{A,12} & 
\lambda \Pi _{A,12}^{\dagger }+\Pi _{A,12}^{\dagger }\Pi _{A,22} \\ 
\lambda \Pi _{A,12}+\Pi _{A,22}\Pi _{A,12} & \Pi _{A,12}\Pi _{A,12}^{\dagger
}+\Pi _{A,22}^{2}%
\end{array}%
\right]  \\
&=&A^{-1}\Pi A
\end{eqnarray*}%
implies%
\begin{eqnarray*}
\left( \lambda ^{2}-\lambda \right) \mathbf{I}_{\mathcal{W}}+\Pi
_{A,12}^{\dagger }\Pi _{A,12} &=&0 \\
\left( \lambda -1\right) \Pi _{A,12}+\Pi _{A,22}\Pi _{A,12} &=&0 \\
\Pi _{A,12}\,\Pi _{A,12}^{\dagger }+\Pi _{A,22}^{2} &=&\Pi _{A,22}.
\end{eqnarray*}%
Define also%
\begin{eqnarray*}
\left\vert succ\right\rangle  &:&=\frac{1}{\sqrt{\lambda }}\Pi A\left\vert
\psi \right\rangle \left\vert 0_{\mathcal{X}}\right\rangle =\frac{1}{\sqrt{%
\lambda }}A\left[ 
\begin{array}{c}
\lambda \left\vert \psi \right\rangle  \\ 
\Pi _{A,12}\left\vert \psi \right\rangle 
\end{array}%
\right] , \\
\left\vert fail\right\rangle  &:&=\frac{1}{\sqrt{1-\lambda }}\left( \mathbf{I%
}-\Pi \right) A\left\vert \psi \right\rangle \left\vert 0_{\mathcal{X}%
}\right\rangle =\frac{1}{\sqrt{1-\lambda }}A\left[ 
\begin{array}{c}
\left( 1-\lambda \right) \left\vert \psi \right\rangle  \\ 
-\Pi _{A,12}\left\vert \psi \right\rangle 
\end{array}%
\right] .
\end{eqnarray*}%
Then we have%
\begin{eqnarray*}
&&AS_{0}^{\phi }A^{-1}S_{1}^{\varphi }\left\vert succ\right\rangle
=AS_{0}^{\phi }A^{-1}S_{1}^{\varphi }A\cdot A^{-1}\left\vert
succ\right\rangle  \\
&=&\frac{A}{\sqrt{\lambda }}\left[ 
\begin{array}{cc}
\phi \left\{ \lambda \left( \varphi -1\right) +1\right\} \mathbf{I}_{%
\mathcal{W}} & \phi (\varphi -1)\Pi _{A,12}^{\dagger } \\ 
(\varphi -1)\Pi _{A,12} & \left( \varphi -1\right) \Pi _{A,22}+\mathbf{I}%
\end{array}%
\right] \left[ 
\begin{array}{c}
\lambda \left\vert \psi \right\rangle  \\ 
\Pi _{A,12}\left\vert \psi \right\rangle 
\end{array}%
\right]  \\
&=&\frac{A}{\sqrt{\lambda }}\left[ 
\begin{array}{c}
\left[ \lambda \phi \left\{ \lambda \left( \varphi -1\right) +1\right\}
+\phi (\varphi -1)\Pi _{A,12}^{\dagger }\,\Pi _{A,12}\right] \left\vert \psi
\right\rangle  \\ 
\left[ \lambda (\varphi -1)\Pi _{A,12}+\left( \varphi -1\right) \Pi
_{A,22}\,\Pi _{A,12}+\Pi _{A,12}\right] \left\vert \psi \right\rangle 
\end{array}%
\right]  \\
&=&\frac{A}{\sqrt{\lambda }}\left[ 
\begin{array}{c}
\left[ \lambda \phi \left\{ \lambda \left( \varphi -1\right) +1\right\}
-\left( \lambda ^{2}-\lambda \right) \phi (\varphi -1)\right] \left\vert
\psi \right\rangle  \\ 
\left[ \lambda (\varphi -1)-\left( \lambda -1\right) \left( \varphi
-1\right) +1\right] \Pi _{A,12}\left\vert \psi \right\rangle 
\end{array}%
\right]  \\
&=&\frac{A}{\sqrt{\lambda }}\left[ 
\begin{array}{c}
\lambda \phi \varphi \left\vert \psi \right\rangle  \\ 
\varphi \Pi _{A,12}\left\vert \psi \right\rangle 
\end{array}%
\right] =A\left[ 
\begin{array}{c}
\sqrt{\lambda }\phi \varphi \left\vert \psi \right\rangle  \\ 
\frac{\varphi }{\sqrt{\lambda }}\Pi _{A,12}\left\vert \psi \right\rangle 
\end{array}%
\right]  \\
&=&\varphi \left( \lambda \phi +1-\lambda \right) \left\vert
succ\right\rangle -\varphi \sqrt{\lambda \left( 1-\lambda \right) }\left(
1-\phi \right) \left\vert fail\right\rangle 
\end{eqnarray*}%
\begin{eqnarray*}
&&AS_{0}^{\phi }A^{-1}S_{1}^{\varphi }\left\vert fail\right\rangle
=AS_{0}^{\phi }A^{-1}S_{1}^{\varphi }A\cdot A^{-1}\left\vert
fail\right\rangle  \\
&=&\frac{A}{\sqrt{1-\lambda }}\left[ 
\begin{array}{cc}
\phi \left\{ \lambda \left( \varphi -1\right) +1\right\} \mathbf{I}_{%
\mathcal{W}} & \phi (\varphi -1)P_{A,12}^{\dagger } \\ 
(\varphi -1)P_{A,12} & \left( \varphi -1\right) P_{A,22}+\mathbf{I}%
\end{array}%
\right] \left[ 
\begin{array}{c}
\left( 1-\lambda \right) \left\vert \psi \right\rangle  \\ 
-P_{A,12}\left\vert \psi \right\rangle 
\end{array}%
\right]  \\
&=&\frac{A}{\sqrt{1-\lambda }}\left[ 
\begin{array}{c}
\left[ \left( 1-\lambda \right) \phi \left\{ \lambda \left( \varphi
-1\right) +1\right\} -\phi (\varphi -1)P_{A,12}^{\dagger }P_{A,12}\right]
\left\vert \psi \right\rangle  \\ 
\left[ \left( 1-\lambda \right) (\varphi -1)P_{A,12}-\left( \varphi
-1\right) P_{A,22}P_{A,12}-P_{A,12}\right] \left\vert \psi \right\rangle 
\end{array}%
\right]  \\
&=&\frac{A}{\sqrt{1-\lambda }}\left[ 
\begin{array}{c}
\left[ \left( 1-\lambda \right) \phi \left\{ \lambda \left( \varphi
-1\right) +1\right\} +\left( \lambda ^{2}-\lambda \right) \phi (\varphi -1)%
\right] \left\vert \psi \right\rangle  \\ 
\left[ \left( 1-\lambda \right) (\varphi -1)+\left( \lambda -1\right) \left(
\varphi -1\right) -1\right] P_{A,12}\left\vert \psi \right\rangle 
\end{array}%
\right]  \\
&=&A\left[ 
\begin{array}{c}
\sqrt{1-\lambda }\phi \left\vert \psi \right\rangle  \\ 
-\frac{1}{\sqrt{1-\lambda }}\Pi _{A,12}\left\vert \psi \right\rangle 
\end{array}%
\right]  \\
&=&-\sqrt{\lambda \left( 1-\lambda \right) }\left( 1-\phi \right) \left\vert
succ\right\rangle +\left( \lambda +\left( 1-\lambda \right) \phi \right)
\left\vert fail\right\rangle .
\end{eqnarray*}%
Therefore, the linear space spanned by $\left\{ \left\vert succ\right\rangle
,\left\vert fail\right\rangle \right\} $ is invariant by the action of $%
AS_{0}^{\phi }A^{-1}S_{1}^{\varphi }$. Especially, in $\phi =\varphi =-1$
case,%
\begin{eqnarray*}
-AS_{0}^{\phi }A^{-1}S_{1}^{\varphi }\left\vert succ\right\rangle  &=&\left(
1-2\lambda \right) \left\vert succ\right\rangle -2\sqrt{\lambda \left(
1-\lambda \right) }\left\vert fail\right\rangle  \\
-AS_{0}^{\phi }A^{-1}S_{1}^{\varphi }\left\vert fail\right\rangle  &=&2\sqrt{%
\lambda \left( 1-\lambda \right) }\left\vert succ\right\rangle +\left(
1-2\lambda \right) \left\vert fail\right\rangle 
\end{eqnarray*}%
and $-AS_{0}^{\phi }A^{-1}S_{1}^{\varphi }$ corresponds to one step of
Grover's search. Therefore, trivially, the repetition of the our
amplification works in the same manner as the case where the auxiliary input
is absent. Also, by choosing the phase factors property, we can control the
speed of the amplification as in \cite{BHT98}.

\subsection{Computational zero-knowledge proof systems for NP}

As is mentioned in subsection 4.2 in \cite{Watrous2005}, a zero-knowledge
proof system for Graph 3-Coloring (G3C) yields a zero-knowledge proof for
any problem in NP. \cite{Watrous2005} presents a simulator for a classical
proof system which is secure against attack by any quantum verifier. In this
subsection, we present a new construction of simulator for this proof system.

In the construction of \cite{Watrous2005}, the essential part is the
amplification of the success probability of a simulator $A$ which succeeds
with probability $\frac{1}{m}$ with $m$ being a polynomially-bounded
function of the input length $n$.

We can construct such an amplification using Grover's amplitude
amplification as is studied in the previous subsection.

On the other hand, the amplification used in \cite{Watrous2005} can be
described in the language of Grover's amplitude amplification as follows.
First, apply $A$ to the initial state $|\psi \rangle |0_{\mathcal{X}}\rangle 
$ , and apply the measurement $\Pi $. If the success event is observed, the
simulation will be successful, and this success event occurs with the
probability $\frac{1}{m}$. Otherwise, the state collapses to $\left\vert
fail\right\rangle $, at which point the reflection operator $%
AS_{0}^{-1}A^{-1}$ is applied. This changes the state to 
\[
\sqrt{\frac{2}{m}}\left\vert succ\right\rangle +\sqrt{1-\frac{2}{m}}%
\left\vert fail\right\rangle , 
\]%
and the measurement $\Pi $ is applied to this, producing $\left\vert
succ\right\rangle $ with the probability $\frac{2}{m}$. The process
continues in this way, with each iteration yielding a successful simulation
with probability at least $\frac{1}{m}$.

\end{document}